\documentclass[         %  
aps,                    %  American Physical Society
prd,                    %  Physical Review D
showpacs,               %  Displays PACS after abstract
nofootinbib,            %  Does not treat footnotes as references
showkeys,               %
preprintnumbers,        %
floatfix]               %  Fixes float errors
{revtex4}               %  REVTEX 4 Package used
\usepackage{graphicx,longtable} 
\begin{document}
%-----------------------------------------------------------------%
\title{Global Analysis of Solar Neutrino and KamLAND Data}
%-----------------------------------------------------------------%
\author{        A.~B. Balantekin}
\email{         baha@nucth.physics.wisc.edu}
%-----------------------------------------------------------------%
\author{        H. Y\"{u}ksel}
\email{         yuksel@nucth.physics.wisc.edu}
%-----------------------------------------------------------------%
\affiliation{  Department of Physics, University of Wisconsin\\
               Madison, Wisconsin 53706 USA }
%-----------------------------------------------------------------%
\date{\today}
%-----------------------------------------------------------------%
\begin{abstract}
A global analysis of the data from all the solar neutrino experiments 
combined with the recent KamLAND data is presented. 
A formula frequently used in the literature 
gives survival probability for three active solar neutrino flavors 
in terms of a suitably-modified two-flavor survival probability. 
Corrections to this formula, which depend on $\theta_{13}$ and 
$\delta m_{31}^2$, 
are calculated. For the mass scale suggested by the atmospheric 
neutrino experiments the contributions of $\delta m_{31}^2$ to these 
corrections is found to be negligible. 
The role of $\theta_{13}$ 
in solar neutrino physics is elaborated. 
For electron 
neutrino oscillations into another active flavor, we find best fit 
values of $\tan^2 \theta_{12} \sim 0.46$, $\tan^2 \theta_{13} \sim 
0$, and $\delta m_{21}^2 \sim 7.1 \times 10^{-5}$ eV$^2$.
It is found that the combined 
solar neutrino and KamLAND data provide the limit $\cos^4 \theta_{13} 
> 0.8$ at the 90 \% confidence level.  
\end{abstract}
%-----------------------------------------------------------------%
\medskip
\preprint{}
\maketitle
%------------------------------------------------------------------%

%-----------------------------------------------------------------%
\section{Introduction}
%-----------------------------------------------------------------%

Seminal developments in neutrino physics took place during the last
few years. Observation of the charged-current solar neutrino flux at
the Sudbury Neutrino Observatory (SNO) \cite{Ahmad:2001an} together
with the measurements of the $\nu_{\odot}$-electron elastic scattering
at the SuperKamiokande (SK) detector \cite{Fukuda:2001nj} established
that there are 
at least two active flavors of neutrinos of solar origin reaching
Earth. Furthermore this SNO charged-current, and subsequent
neutral-current \cite{Ahmad:2002jz}  and day-night difference
\cite{Ahmad:2002ka} measurements confirmed the prediction of the
Standard Solar Model (SSM) for the total neutrino flux
\cite{Bahcall:2000nu}. Already in
the summer of 2002 global analyses combining SNO and
SuperKamiokande data with earlier chlorine \cite{Cleveland:nv} and
gallium \cite{Abdurashitov:2002nt,Hampel:1998xg,Altmann:2000ft}
measurements indicated the so-called large mixing angle (LMA) region
of the neutrino parameter space as the most likely solution  
\cite{Bahcall:2002hv}.

In December 2002 the KamLAND reactor neutrino experiment announced its
first results \cite{:2002dm}. The KamLAND data also points to  the LMA
region, a result which was confirmed by independent analyses
\cite{Bahcall:2002ij,Fogli:2002au,Barger:2002at,Maltoni:2002aw,%
Bandyopadhyay:2002en,Nunokawa:2002mq,Aliani:2002na,deHolanda:2002iv,%
Creminelli:2001ij}. 
One aim of our paper is to present a global analysis of 
all solar neutrino data along with the recent KamLAND data 
to examine neutrino parameter space. Our other goal is elaborate on 
the role of the mixing angle between the first and the third 
neutrino families, $\theta_{13}$, in solar neutrino physics. 
Currently the value of $\theta_{13}$ is probably the most-pressing open 
question in neutrino physics. Since the CP-violating phase appears 
together with $\theta_{13}$, understanding CP violation in 
neutrino oscillations 
\cite{Beavis:2002ye,Minakata:2002qi,Akhmedov:2001kd}
requires precise knowledge of $\theta_{13}$. A priori one expects 
the impact of $\theta_{13}$ on solar neutrino physics to be minimal. 
However after recent significant accumulation of neutrino data we 
can now provide a more precise assessment of this conjecture. 

A formula frequently used in the literature 
gives survival probability for three active solar neutrino flavors 
in terms of a suitably-modified two-flavor survival probability. We 
calculate systematic corrections to this formula in Section II.
These corrections depend on $\theta_{13}$ and 
$\delta m_{31}^2$, however or the mass scale indicated by the 
atmospheric 
neutrino experiments the contributions of $\delta m_{31}^2$ to these 
corrections is found to be negligible.     
We use this formula to investigate the role of  $\theta_{13}$ in 
solar neutrino physics. 
Details of our statistical analysis are presented in Section III.
Finally Section IV contains a discussion of our results.

%-------------------------------------------------------------%
\section{Matter-Enhanced Neutrino Oscillations}
%----------------------------------------------------------------%

We denote the neutrino mixing
matrix by $U_{\alpha i}$ where $\alpha$ denotes the
flavor index and $i$ denotes the mass index:
\begin{equation}
  \label{udef}
  \Psi_{\alpha} = \sum_i U_{\alpha i} \Psi_i . 
\end{equation}
For three neutrinos we take
\begin{equation}
\label{mixing}
U_{\alpha i} = T_{23}T_{13}T_{12} = \left(\matrix{
     1 & 0 & 0  \cr
     0 &  C_{23}  & S_{23} \cr
     0 & - S_{23} &  C_{23} }\right)
 \left(\matrix{
     C_{13} & 0 &  S_{13}^{\ast} \cr
     0 &  1 & 0 \cr
     - S_{13} & 0&  C_{13} }\right)
 \left(\matrix{
     C_{12} & S_{12} &0 \cr
     - S_{12} & C_{12} & 0 \cr
     0 & 0&  1 }\right) ,
\end{equation}
where $C_{13}$, etc. is the short-hand notation for $\cos
{\theta_{13}}$, etc. Note that individual matrices, not their matrix
elements, are called $T_{23}$, $T_{13}$, and $T_{12}$, respectively in
Eq. (\ref{mixing}) . In $T_{13}$
the notation $S_{13}^{\ast}$ was used to indicate
$(\sin{\theta_{13}})e^{i\phi}$ where $\phi$ is the CP-violating
phase. We will ignore this phase in our discussion.                 
The evolution of the three neutrino species in matter is governed by
the MSW equation \cite{Wolfenstein:1977ue,Mikheev:gs,Mikheev:wj}: 
\begin{equation}
\label{msw}
i \frac{\partial}{\partial t} \left(\matrix{ \Psi_e\cr \Psi_{\mu} \cr
    \Psi_{\tau}} \right)  = \left[ T_{23}T_{13}T_{12} \left(\matrix{
     E_1 & 0 & 0  \cr
     0 &  E_2  & 0 \cr
     0 & 0 &  E_3 }\right)
T_{12}^{\dagger}T_{13}^{\dagger} T_{23}^{\dagger} +
\left(\matrix{
     V_c+V_n & 0 & 0  \cr
     0 &  V_n  & 0 \cr
     0 & 0 &  V_n }\right)\right] \left(\matrix{ \Psi_e\cr \Psi_{\mu}
    \cr     \Psi_{\tau}} \right),
\end{equation}
where the Wolfenstein potentials in neutral, unpolarized matter are
given by 
\begin{equation}
  \label{wolfen1}
V_c (x) = \sqrt{2} G_F  N_e (x) 
\end{equation}
for the charged-current and 
\begin{equation}
  \label{wolfen2}
V_n (x) =  - \frac{1}{\sqrt{2}} G_F N_n(x).
\end{equation}
for the neutral current. Since $V_n$ only contributes an overall phase
to the neutrino evolution in the Sun we will ignore it in the rest 
of this paper. 

Following Ref. \cite{Balantekin:1999dx} we introduce the
combinations 
\begin{eqnarray}
\label{rot1}
\tilde{\Psi}_{\mu} &=& \cos{\theta_{23}} \Psi_{\mu} -
\sin{\theta_{23}} \Psi_{\tau}, \\
\tilde{\Psi}_{\tau} &=& \sin{\theta_{23}} \Psi_{\mu} +
\cos{\theta_{23}} \Psi_{\tau},
\end{eqnarray}
which corresponds to multiplying the neutrino column vector in
Eq. (\ref{msw}) with $T_{23}^{\dagger}$ from the left. Eq. (\ref{msw})
now takes the form
\begin{equation}
\label{mswmod1}
i \frac{\partial}{\partial t} \left(\matrix{ \Psi_e\cr
    \tilde{\Psi}_{\mu} \cr \tilde{\Psi}_{\tau} }\right)  
= \left[       T_{13}T_{12} \left(\matrix{
     E_1 & 0 & 0  \cr
     0 &  E_2  & 0 \cr
     0 & 0 &  E_3 }\right)
T_{12}^{\dagger}T_{13}^{\dagger}                  +
\left(\matrix{
     V_c   & 0 & 0  \cr
     0 &  0    & 0 \cr
     0 & 0 &  0   }\right)\right] \left(\matrix{ \Psi_e\cr
    \tilde{\Psi}_{\mu}  \cr     \tilde{\Psi}_{\tau} }\right),
\end{equation}
where we used the the fact that $T_{23}$ commutes with the second
matrix on the right containing the matter potential. Note that the
initial conditions on $\tilde{\Psi}_{\mu}$ and $\tilde{\Psi}_{\tau}$
are the same as those on $\Psi_{\mu}$ and $ \Psi_{\tau}$: They are 
initially all zero. Consequently the mixing angle $\theta_{23}$ does
not enter in the source- and detector-averaged solutions of the
matter-enhanced neutrino oscillations. 

We next perform the transformation
\begin{equation}
\label{rot2}
 \left(\matrix{ \varphi_e\cr
    \varphi_{\mu} \cr \varphi_{\tau} }\right) = T_{13}^{\dagger}
 \left(\matrix{ \Psi_e\cr
    \tilde{\Psi}_{\mu} \cr \tilde{\Psi}_{\tau} }\right) = 
 \left(\matrix{\cos{\theta_{13}} \Psi_e + \sin{\theta_{13}}
    \tilde{\Psi}_{\tau} \cr
    \tilde{\Psi}_{\mu} \cr - \sin{\theta_{13}}  \Psi_e +
    \cos{\theta_{13}} \tilde{\Psi}_{\tau} }\right),
\end{equation}
after which Eq. (\ref{mswmod1}) takes the form
\begin{equation}
\label{mswmod2}
i \frac{\partial}{\partial t} \left(\matrix{ \varphi_e\cr
    \varphi_{\mu} \cr \varphi_{\tau} }\right) = {\cal H} 
\left(\matrix{ \varphi_e\cr
    \varphi_{\mu} \cr \varphi_{\tau} }\right),
\end{equation}
where we dropped a term proportional to the identity. In this
equation ${\cal H}$ is given by 
\begin{equation}
\label{h0}
{\cal H} = \left(\matrix{
     \frac{1}{2} \tilde{V} - \Delta_{21} \cos 2 \theta_{12}&
     \frac{1}{2} \Delta_{21} \sin 2 \theta_{12} & - \frac{1}{2} V_c
     \sin 2 \theta_{13} \cr
     \frac{1}{2} \Delta_{21} \sin 2 \theta_{12} & -
     \frac{1}{2}\tilde{V} + \Delta_{21} \cos 2 \theta_{12}   & 0 \cr
     - \frac{1}{2} V_c \sin 2 \theta_{13}  & 0 &
     \frac{1}{2}(\Delta_{31}+\Delta_{32}) + V_c - \frac{3}{2} 
     \tilde{V} }\right),
\end{equation}
where we introduced the modified matter potential 
\begin{equation}
\label{modpotent}
\tilde{V} = V_c\cos^2\theta_{13}
\end{equation}
and the quantity 
\begin{equation}
\label{Delta}
\Delta_{ij} = \frac{m_i^2 - m_j^2}{2E} = \frac{\delta m_{ij}^2}{2E}. 
\end{equation}
The corresponding mass matrix was considered in
Ref. \cite{Kuo:1986sk}. 

Eq. (\ref{mswmod2}) describes the exact evolution of the neutrino
eigenstates through matter. Assuming the mass hierarchy $m_3 > m_2 >
m_1$, there are two MSW resonances. The lower-density resonance is
apparent in this equation. Although it is not immediately obvious this
equation also correctly describes the higher density resonance. 
One observes from Eq. (\ref{rot2}) that the appropriate initial
conditions are            
\begin{equation}
\label{init2}
\left(\matrix{ \varphi_e (t=0)\cr
    \varphi_{\mu} (t=0)\cr \varphi_{\tau} (t=0)}\right) = 
\left(\matrix{\cos{\theta_{13}}  \cr
    0 \cr  - \sin{\theta_{13}} }\right) 
\end{equation}
These initial conditions need to be satisfied where the neutrino is
produced, which is not necessarily at the center of the Sun. 

We look for the solution of Eq. (\ref{mswmod2}) appropriate for the
solar neutrino problem. Specifically we assume that
$\sin{\theta_{13}}$ is small as indicated by the reactor neutrino
experiments prior to  KamLAND
\cite{Bemporad:2001qy,Apollonio:1999ae,Boehm:2001ik}. We also
interpret the value $ \delta m_{32}^2 = 3 \times 10^{-3}$ eV$^2$
observed by the atmospheric neutrino observations at the
SuperKamiokande detector \cite{Fukuda:2000np} to imply that $ \delta
m_{32}^2 \sim \delta m_{31}^2$ since the LMA value of $\delta
m_{21}^2$ deduced from the solar neutrino experiments is lower by more
than one order of magnitude. Hence we take $\Delta_{31} \sim
\Delta_{32}$. Using the approximate electron density given by Bahcall
\cite{Bahcall:ks} 
\begin{equation}
\label{appelecden}
N_e (r) = 245 \>{\rm exp}(- 10.54 r/ R_{\odot}) N_A cm^{-3},
\end{equation}
where $N_A$ is the Avogadro's number and $R_{\odot}$ is the radius of 
the Sun, we write the potential of Eq. (\ref{wolfen1}) in the
appropriate units as 
\begin{equation}
\label{wolfen1prime}
V_c (r) = 1.87 \times 10^{-5} \times {\rm exp}(- 10.54 r/ R_{\odot})
\>\> {\rm eV}^2/{\rm MeV}. 
\end{equation}
Using the value $ \delta m_{31}^2 \sim 10^{-3}$  eV$^2$ and the
Eq. (\ref{wolfen1prime}), even for the highest energy ($\sim 15$ MeV)
solar neutrinos the condition 
\begin{equation}
\label{pertcond}
\frac{V_c (r)}{\Delta_{31}} < 1
\end{equation} 
is satisfied everywhere in the Sun. Indeed as one moves away from the 
center of the Sun this ratio is much smaller especially for the 
lower-energy solar neutrinos. We will use this ratio as a perturbation
parameter in solving the neutrino propagation equations. 

The differential equation involving the derivative of $\varphi_{\tau}$ 
is 
\begin{equation}
\label{taueq}
i \frac{\partial}{\partial t} \varphi_{\tau} (t) = a \varphi_e + b
\varphi_{\tau}, 
\end{equation} 
where we introduced the abbreviated notation
\begin{equation}
\label{a}
a = - V_c \sin{\theta_{13}} \cos{\theta_{13}}
\end{equation}          
and 
\begin{equation}
\label{b} 
b = \frac{1}{2}(\Delta_{31}+\Delta_{32}) + V_c - \frac{3}{2} \tilde{V}
\simeq \Delta_{31} - \frac{1}{2} V_c (1 - 3 \sin^2{\theta_{13}}) 
\equiv \Delta_{31} - \epsilon . 
\end{equation} 
Using the initial conditions of Eq. (\ref{init2}), Eq. (\ref{taueq})
can be immediately solved  to express $\varphi_{\tau}$ in terms of
$\varphi_e$: 
\begin{equation}
\label{phitau}
\varphi_{\tau} (t) = \sin{\theta_{13}} \> e^{-i\int_{0}^{t} b (t')
  dt'} 
\left[ - 1 + i \cos{\theta_{13}} \int_{0}^{t} dt' V_c (t')
  \varphi_e(t') 
e^{i\int_{0}^{t'} b (t'') dt''}  \right].
\end{equation} 
Of course in general one still needs the value of $\varphi_e$
everywhere 
in the Sun. However when the limiting condition of
Eq. (\ref{pertcond}) is satisfied one can use a technique employed in
Ref. \cite{Balantekin:1984jv} to obtain an approximate
expression. Writing $ d e^{i \Delta_{31} t} = i \Delta_{31} e^{i
  \Delta_{31} t} dt$ one can integrate the integral in
Eq. (\ref{phitau}) by parts once to obtain
\begin{equation}
\label{firstit}
i \int_{0}^{t} dt' V_c  \varphi_e e^{i\int_{0}^{t'} b (t'') dt''}
= - \frac{V_c(t=0)}{\Delta_{31}} \cos{\theta_{13}} -
\frac{1}{\Delta_{31}} \int_0^t e^{i \Delta_{31} t'} \frac{d}{dt'}
\left[ 
  V_c(t')\varphi_e(t') e^{i\int_{0}^{t'} \epsilon (t'') dt''} \right]
\end{equation}
where we used the initial condition of Eq. (\ref{init2}) on
$\varphi_e$ 
and the fact that $V_c$ is zero once the neutrinos leave the Sun. In
the last integral in Eq. (\ref{firstit}) we ignore the derivative of
$V_c$ since it is many orders of magnitude smaller than $V_c$ itself
at the core of the Sun (cf. Eq. (\ref{wolfen1prime})). We calculate
the derivative of $\varphi_e$ using the
Eq. (\ref{mswmod2}). Integrating 
this last integral by parts again we obtain an expansion of
$\varphi_{\tau}$ {\em outside the Sun}:
\begin{eqnarray}
\label{phitau2}
&\varphi_{\tau}& (t) = \sin{\theta_{13}} \> e^{-i\int_{0}^{t} b (t') 
  dt'} 
\nonumber \\ &\times& 
\left\{ -1 + \cos^2{\theta_{13}} \left[ -
\xi \left( 1 + \frac{\delta m_{21}^2}{\delta
    m_{31}^2} + \cdots \right) + 
\xi^2 
\left( 1 - 2 \sin^2{\theta_{13}} + \cdots \right) \right]
\right\}
\end{eqnarray} 
where we defined
\begin{equation}
\label{v0def}
\xi \equiv \frac{V_c(t=0)}{\Delta_{31}}.
\end{equation}
Note that Eq. (\ref{phitau2}) does not represent $\varphi_{\tau}$ inside
the Sun since we have already taken the limit $V_c=0$. 

The differential equation one can write from Eq. (\ref{mswmod2})
involving the derivative of $\varphi_e$ includes a term containing
$\varphi_{\tau}$ ( $- \frac{1}{2} V_c \sin 2
\theta_{13}\varphi_{\tau}$). However substitution of  $\varphi_{\tau}$
into 
this term using Eq. (\ref{phitau2}) indicates that this term is second
order in $\sin{\theta_{13}}$ and can be discarded in a calculation
which is first order in $\sin{\theta_{13}}$. In this approximation the
evolution of the first two components in Eq. (\ref{mswmod2}) decouple
from the third yielding
\begin{equation}
\label{mswmod3}
i \frac{\partial}{\partial t} \left(\matrix{ \varphi_e\cr
    \varphi_{\mu} } \right) = 
\left(\matrix{
     \frac{1}{2} \tilde{V} - \Delta_{21} \cos 2 \theta_{12}&
     \frac{1}{2} \Delta_{21} \sin 2 \theta_{12} \cr
     \frac{1}{2} \Delta_{21} \sin 2 \theta_{12} & -
     \frac{1}{2}\tilde{V} + \Delta_{21} \cos 2 \theta_{12} }\right) 
\left(\matrix{ \varphi_e \cr \varphi_{\mu} }\right),
\end{equation}
which is the standard MSW evolution equation for two neutrinos except
that the electron density $N_e$ is replaced by $N_e
\cos^2\theta_{13}$. The fact that one needs to solve the evolution
equations with this new density was already emphasized in the
literature \cite{Fogli:1996ne,Fogli:2000bk}. Since this $2\times 2$
``Hamiltonian'' in Eq. (\ref{mswmod2}) is an element of the SU(2)
algebra, the resulting evolution operator is an element of the SU(2)
group \cite{Balantekin:1998yb} and the evolving states can be written
as
\begin{equation}
\label{evolve}
\left(\matrix{ \varphi_e (t)\cr \varphi_{\mu} (t) } \right) = 
\left(\matrix{
     \Phi_e (t)&  - \Phi_{\mu}^*(t) \cr
     \Phi_{\mu}(t) & \Phi_e^* (t)}\right) 
\left(\matrix{ \cos{\theta_{13}} \cr 0 }\right)
\equiv {\hat U}\left(\matrix{ \cos{\theta_{13}} \cr 0 }\right) ,
\end{equation}
where $\Phi_e(t)$ and $\Phi_{\mu}(t)$ are solutions of Eq. 
(\ref{mswmod2}) with the initial conditions  $\Phi_e (t=0) = 1$ 
and $\Phi_{\mu} (t=0) = 0$.

We can write the electron neutrino amplitude using Eq. (\ref{rot2}) as
\begin{equation}
\label{finalnue}
\Psi_e = \cos{\theta_{13}} \> \varphi_e - \sin{\theta_{13}} \>
\varphi_{\tau}. 
\end{equation}
Substituting $\varphi_e$ from Eq. (\ref{evolve}) and $\varphi_{\tau}$
from 
Eq. (\ref{phitau2}) into Eq. (\ref{finalnue}) we obtain
\begin{equation}
\label{finalnue1}
\Psi_e = \cos^2{\theta_{13}} \> \Phi_e - \sin^2{\theta_{13}}
e^{-i\int_{0}^{t} b (t') dt'} D,
\end{equation}
where we defined $D \equiv \varphi_{\tau}e^{i\int_{0}^{t} b (t') dt'}
/ 
\sin{\theta_{13}}$ for convenience of notation. Squaring
Eq. (\ref{finalnue1}) we finally obtain the electron neutrino survival
probability in our approximation:
\begin{eqnarray}
\label{finalnue2}
&P&_{3\times3}( \nu_e \rightarrow  \nu_e) = \cos^4{\theta_{13}} \> 
P_{2\times2}( \nu_e \rightarrow  \nu_e \>{\rm with}\> N_e
\cos^2{\theta_{13}}) \nonumber \\
&+& \sin^4{\theta_{13}} \left[ 1 + 2 \xi \cos^2{\theta_{13}}  \left(
    1 + \frac{\delta m_{21}^2}{\delta m_{31}^2} \cos 2 \theta_{12}
  \right) + \xi^2  \cos^4{\theta_{13}} \left( 
     2 \frac{\delta m_{21}^2}{\delta m_{31}^2} \cos 2 \theta_{12}
  - 1 \right) \right] \nonumber\\
&+& {\cal O} (\xi^3) .
\end{eqnarray}
$\xi$ in this equation is given by Eq. (\ref{v0def}). $P_{2\times2}(
\nu_e \rightarrow  \nu_e \>{\rm with}\> N_e \cos^2{\theta_{13}})$ is
the standard 2-flavor survival probability calculated with the
modified electron density $N_e \cos^2{\theta_{13}}$ and the standard
initial 
conditions ($\Phi_e=1$ and $\Phi_{\mu}=0$). 
Either a number of exact
\cite{Petcov:1987zj,Notzold:1987cq,Bruggen:hr,Balantekin:1997jp,%
Lehmann:2000ey}
or approximate 
\cite{Balantekin:1996ag,Balantekin:1997fr,Lisi:2000su} 
solutions of 
the neutrino evolution equations available in the literature or
numerical methods can be used to calculate the survival probability of
Eq. (\ref {finalnue2}). 

\begin{figure}
\includegraphics[scale=.85, bb= 100 100 510 750]{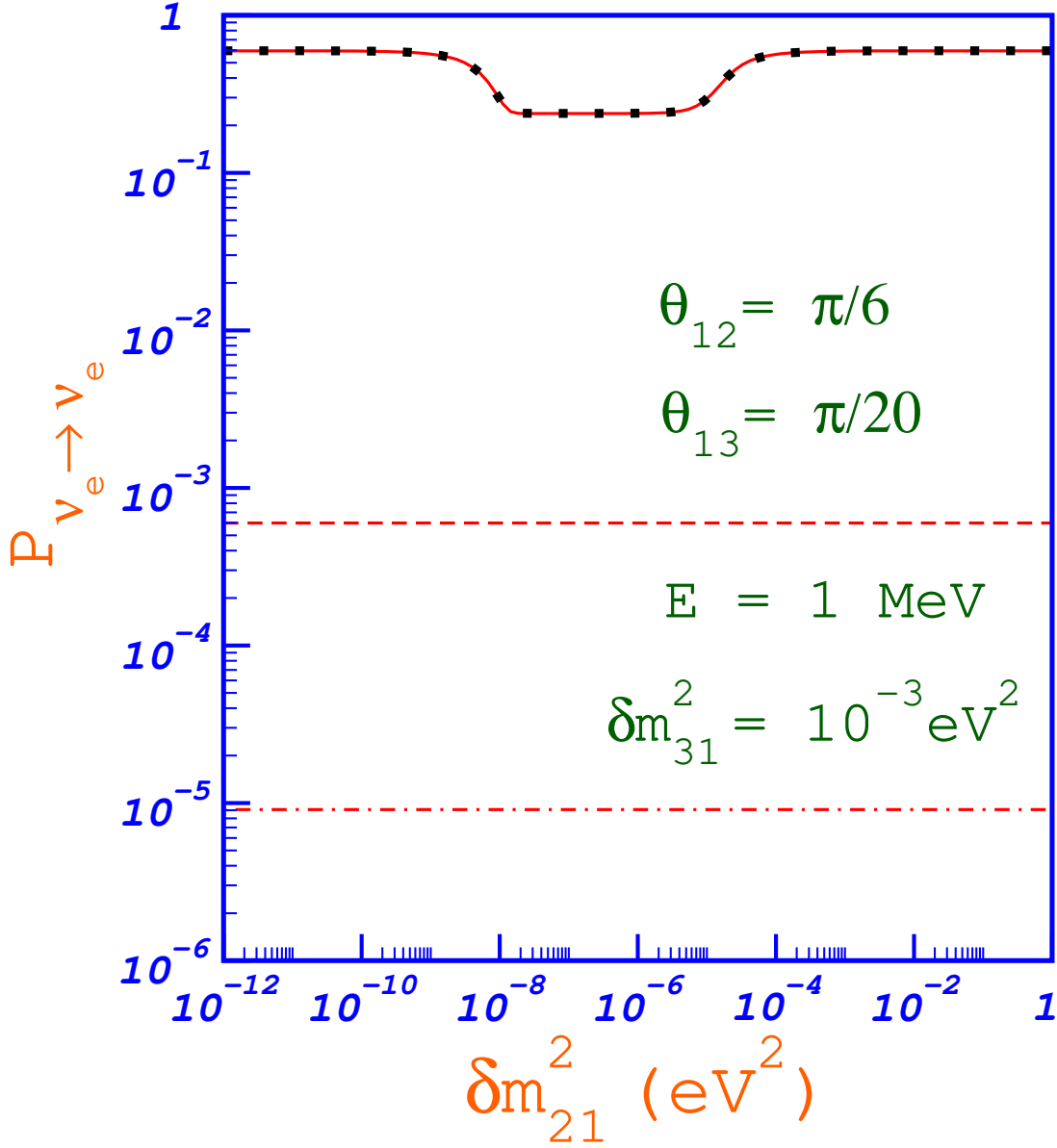}
\vspace*{+2cm} \caption{ \label{fig:1} 
  Several components of the electron neutrino survival
  probability given in Eq. (\ref{finalnue2}). In this figure
  $\theta_{13} = \pi/6$, $\theta_{13}  = \pi/20$, $E_{\nu} = 1$ MeV,
  and $\delta m_{31}^2 = 10^{-3}$ eV$^2$. The upper solid line is the
  total survival probability. The filled squares on top of the figure 
  represent the contribution of the term proportional to 
  $\cos^4{\theta_{13}}$ alone to the survival probability. The middle 
  long-dashed line is the contribution of the term proportional to 
  $\sin^4{\theta_{13}}$ alone to the survival probability. The lower 
  dash-dotted line is the contribution of the correction terms 
  proportional to $\xi$ only.}
\end{figure}

For $\xi=0$
Eq. (\ref{finalnue2}) reduces to the formula widely used in the
literature (see e.g. Refs. \cite{Fogli:2000bk,Kuo:1989qe}). 
We show the relative contributions of various terms to the survival
probability given in Eq. (\ref{finalnue2}) in Fig. \ref{fig:1}.
We picked values of the neutrino parameters representative of the 
solar neutrino results in this figure. The upper solid line is the 
total survival probability numerically calculated for the 3-flavor 
mixing by solving the neutrino evolution equation, Eq. (\ref{msw}), 
exactly using the method of Ref. \cite{Ohlsson:1999um}.   
The filled squares on top of the figure represent the 
contribution of the term proportional to $\cos^4{\theta_{13}}$ alone 
(the first term in the right side of Eq. (\ref{finalnue2})) to the 
survival probability in the approximation described above. One 
observes that this term alone is an excellent approximation to the 
total probability. To illustrate this we plot separately the 
contribution of the term proportional to $\sin^4{\theta_{13}}$ alone 
to the survival probability (the long-dashed line in the middle). 
Clearly the second term is approximately three orders of magnitude 
smaller than the first one. The correction terms proportional to 
the parameter $\xi$ are even smaller by two more orders of 
magnitude indicating the validity of our expansion in terms of the 
parameter $\xi$. It is worth emphasizing that the quantity 
$\delta m_{31}^2$ enters the survival probability in Eq. 
(\ref{finalnue2}) only through the terms proportional to $\xi$, 
hence its effect is minimal: For solar neutrinos the 3-flavor 
survival probability depends only on two mixing angles, 
$\theta_{12}$, $\theta_{13}$, and one mass-difference squared, 
$\delta m_{21}^2$. Having gained confidence in the validity of 
Eq. (\ref{finalnue2}), we use it in our analysis.

%-----------------------------------------------------------%
\section{Statistical Analysis}
%-----------------------------------------------------------%

There is an extensive literature describing methods to calculate 
the goodness of a fit and confidence levels of allowed regions 
(See e.g. Refs. 
\cite{Feldman:1997qc,Fogli:2002pt,Garzelli:2000yf,%
Gonzalez-Garcia:2002dz}
and other references we cite below).  
In our global analysis, we use the ``covariance approach'', in which
least-squares function for solar data is defined as
\begin{equation}\label{e1}
\chi^{\, 2}_{\, \odot}=\sum_{ij} \,
(R_{\,i}^{\, exp}-R_{\,i}^{\, th}) \, \sigma^{-2}_{\,ij} \,
(R_{\,j}^{\, exp}-R_{\,j}^{\, th})\, , 
\end{equation}
where $R_{\,i}^{\, exp}$ and $R_{\,i}^{\, th}$ are the experimental
values and theoretical predictions of the observables respectively 
and $\sigma^{\,-2}$ 
is the inverse of the covariance error matrix built from the
statistical and systematic errors considering mutual correlations. 

We use 80 data points in our analysis; the  total rate of the chlorine
experiment (Homestake), the average rate of the gallium experiments
(SAGE, GALLEX, GNO),  44 data points from the SK zenith-angle-spectrum
and 34 
data points from the SNO day-night-spectrum.  The only correlation
between the rates of the water Cerenkov experiments and radiochemical
experiments is the uncertainty in the $ ^8B$ flux. Since we fit 
the shape of the spectrum for water Cerenkov experiments in our 
analysis,  the covariance error matrix can be block diagonalized. We
write contributions to $\chi^2_{\,\odot}$ from the rates of the 
radiochemical experiments, the SK zenith-angle-spectrum, and
the SNO day-night-spectrum explicitly:
\begin{equation}\label{e2}
\chi^{\, 2}_{\, \odot}=\chi^{\, 2}_{\, {\rm Cl, Ga \, Rates}}+\chi^{\,
  2}_{\rm \, SK}+\chi^{\, 2}_{\rm \, SNO} .
\end{equation}
SK and SNO uses the pattern and intensity of the Cerenkov light
generated by the recoiling electron in order to detect events due to
 electron scattering (ES):
\begin{eqnarray}\label{e3}
\nu_x + e^- &\to& \nu_x + e^-\   \text{ (ES)} .
\end{eqnarray}
ES is sensitive to all neutrino flavors with reduced sensitivity to
non-electron-neutrino components. Since SNO contains heavy water it is
sensitive to charge-current (CC) and neutral-current (NC)
reactions in addition to ES: 
\begin{eqnarray}\label{e4}
\nu_e + d &\to& p + p + e^-\ ,  \text{ (CC)}, \\
\nu_x + d &\to& n + p + \nu_x   \ , \text{ (NC).}
\end{eqnarray}
The Cerenkov light generated by the recoiling electron is used to
observe the CC 
events while the  gamma ray from the neutron capture on  deuterium is
used to detect NC events. Time, location, direction, and energy of the
CC events allow reconstruction of the solar neutrino spectrum. 

Since Cerenkov experiments have a higher threshold energy they 
are sensitive to only $^8B$ and $hep$ neutrinos. The $hep$ neutrino
flux is  much smaller than the $^8B$ neutrino flux, but  $hep$
neutrinos are somewhat more  energetic. The production regions for
these two components of the neutrino flux are not much different. For
the sake of simplicity instead of dealing with two  different sources
of neutrinos, we add $hep$ neutrino spectrum to the  $^8B$ neutrino
spectrum 
\begin{eqnarray}
\label{e5}
\lambda_{B}(E_\nu) &\rightarrow& \frac{\phi_{B} \lambda_{B}(E_\nu) + 
\phi_{hep}\lambda_{hep}(E_\nu)} {\phi_{B}+\phi_{hep}},\\
\phi_{B} &\rightarrow& \phi_{B} + \phi_{hep},
\end{eqnarray}
and use them as a single source in the analysis of Cerenkov
experiments. 

Cerenkov experiments are also live-time experiments. They can
measure separate day and  night rates or even divide their night rate
into several zenith angle ``$\alpha$''  bins. We  may expect  to see a
different rate for each of these bins since MSW mechanism predicts 
earth
regeneration  effects at night when neutrinos pass through several
layers of Earth material.  We incorporate those effects into our
analysis by calculating the survival probability  numerically at each
zenith angle using a step function density approximation to the
Preliminary Earth Model \cite{Dziewonski:xy}. Survival probability for
each zenith-angle bin is calculated
by averaging the probability weighted with the exposure function
``$f(\alpha)$'' of  the detector. For SK we used exposure function
given 
in Ref. \cite{Bahcall:1997jc}. For SNO we only used day-night bins and
live-time information from Ref. \cite{HOWTOSNO}. For any zenith bin
between $\alpha_{min}^j$ and $\alpha_{max}^j$, the ``weighted'' 
average survival
probability $\langle P^{\,  j} (\nu_e \rightarrow \nu_e, E_\nu)
\rangle$ is:  
\begin{equation}
\label{e6}
\langle P^{\, j} (\nu_e \rightarrow \nu_e, E_\nu) \rangle =
\frac{\int_{\alpha_{min}^j}^{\alpha_{max}^j} f(\alpha)  P^{\,} (\nu_e
  \rightarrow \nu_e, E_\nu,\alpha)  
\, d\alpha}
{ \int_{\alpha_{min}^j}^{\alpha_{max}^j} f(\alpha) \, d\alpha }
\end{equation}

SK measures the kinetic energy of the recoiling electron  and reports
the data divided into several kinetic energy intervals. The kinetic
energy  assigned to the event by the detector is not always same as
the true kinetic energy, instead  it has a distribution around actual
kinetic energy. This is characterized by a Gaussian shaped response
function \cite{Bahcall:1996ha}:
\begin{eqnarray}
\label{e7}
R(T,\,T^\prime) &=& \frac{1}{\sqrt{2\pi}s_{\rm SK}}\exp\left[
{-\frac{(T-T^\prime)^2}{2 s_{\rm SK}^2}}\right]\ ,\\ \label{e7a} 
s_{\rm SK} & = & 0.47 \sqrt{T^\prime} \> \text{MeV}, 
\end{eqnarray}
where $s_{\rm SK}$ is the width of the Gaussian \cite{Faid:1996nx},
$T^\prime$ 
is the actual kinetic energy of the recoiling electron and $T$ is the
kinetic energy assigned to the same event by the detector (in
Eqs. (\ref{e7}) and (\ref{e7a}) both $T$ and $T'$ are in
MeV).  By convolving  differential ES cross sections of
Ref. \cite{Bahcall:1995mm} with  energy response function of
the detector at each kinetic energy interval $i$,  we get the
``corrected'' cross sections corresponding to the kinetic energy bin 
$i$: 
\begin{equation}
\sigma_{\nu_e,\nu_x}^i(E_\nu)= \int_{T_{ min}^i}^{T_{ max}^i}\!dT
\int_0^{T_{max}^\prime}\!dT^\prime\, R (T,\,T^\prime)\,
\frac{d\sigma_{\nu_e,\nu_x}(E_\nu,\,T^\prime)}{dT^\prime}\ ,
\label{corrected1}
\end{equation}
where $ T_{max}^\prime = E_\nu / (1+m_e/2 E_\nu) $ is the maximum
kinetic energy that  any electron can have due to kinematical
limits. After this step we have all the ingredients to calculate the
rates for the zenith-spectrum bins, weighted survival  probabilities
for zenith-angle bins  ($\langle P^{\, j} (\nu_e \rightarrow \nu_e,
E_\nu) \rangle$), and the corrected cross
sections for kinetic energy spectrum bins ($\sigma_{\nu_e,\nu_x}^i
(E_\nu)$).  If we consider only oscillations into active flavors
the theoretical rate at SK for each zenith-spectrum bin is 
\begin{eqnarray}
\label{e8}
R^{SK}_k= R^{SK}_{i,j} &=& \phi_B\int\! dE_\nu\, \lambda_B (E_\nu)
[ \sigma_{\nu_e}^i(E_\nu)  \langle P^j (\nu_e \rightarrow \nu_e,
E_\nu,t) \rangle \nonumber \\ &+& \sigma_{\nu_x}^i(E_\nu) (1-\langle
P^j (\nu_e \rightarrow \nu_e, E_\nu,t)\rangle ) ],
\end{eqnarray} 
where we use ``$k$'' as a collective index for zenith-spectrum bins
instead of the pair ``$i,j$''.  SK reports the ratio of number of 
observed 
events to the number of expected  events under no oscillation
condition. Dividing the above rate with SSM expected
value (i.e. with the survival probability is 1) we obtain ratios  
to be compared with those given in 
Ref. \cite{Fukuda:2002pe}. 

To calculate the error matrix for SK zenith-spectrum bins,
experimental 
rates and  uncertainties are taken from \cite{Fukuda:2002pe}.  
For each zenith and energy
bin SK reports rates and  statistical and systematic uncertainties.  
We take the $^8B$ shape,
energy scale and energy resolution uncertainties from
\cite{Fogli:2002pt} which are
actually calculated under no-oscillation  condition and may result
in discrepancies at higher confidence levels. 
An additional overall systematic offset error of 2.75\%
is added to all bins. 
In calculating $\chi^{\,  2}_{\rm \,SK}$ we introduce a free
normalization 
parameter ``$\eta$'' and minimize $\chi^{\, 2}_{\rm \,SK}$ with 
respect to 
$\eta$. In this way the total $^8B$ flux allowed to float freely.
\begin{equation}\label{e9}
\chi^{\, 2}_{\rm \, SK}=\sum_{ij} \,
(R_{\,i}^{\, exp}- \eta R_{\,i}^{\, th}) \, \sigma^{-2}_{\,ij} \,
(R_{\,j}^{\, exp}-\eta R_{\,j}^{\, th})\, , 
\end{equation}

SNO response
function for electrons is \cite{Ahmad:2002jz}:
\begin{eqnarray}
\label{e10}
R(T,\,T^\prime) &=& \frac{1}{\sqrt{2\pi}s_{\rm SNO}}\exp\left[
{-\frac{(T-T^\prime)^2}{2 s_{\rm SNO}^2}}\right]\ ,\\
s_{\rm SNO} &=& (-0.0684 +0.331\sqrt{T^\prime}+0.0425 T^\prime )
\text{   MeV}, 
\end{eqnarray}
``Corrected'' ES and CC cross sections can be  obtained  in a similar
fashion as we did for SK (cf. Eq. ( \ref{corrected1})). NC events are
mono-energetic. Neutrons are first thermalized and then captured on
deuterium.  All NC events originally have same energy, $T_{NC}=5.08$
MeV. Their response function is \cite{HOWTOSNO}
\begin{eqnarray}
\label{e11}
R(T) &=& \frac{1}{\sqrt{2\pi}s_{\rm NC}}\exp\left[
{-\frac{(T-T_{NC})^2}{2 s^2_{\rm NC}}}\right]\ ,\\
s_{\rm NC} &=& 1.11 \text{MeV}, 
\end{eqnarray}
and the corresponding ``corrected'' cross sections are: 
\begin{equation}
\sigma_{NC}^i(E_\nu)=\int_{T_{ min}^i}^{T_{ max}^i}\!dT
R (T)\,
\frac{d\sigma_{NC}(E_\nu,\,T)}{dT}\ .
\label{e12}
\end{equation}
The response function of Eq. (\ref{e11}) 
spreads the neutral current events in energy. 
We use the ``forward-fitting'' technique described in
\cite{HOWTOSNO} to calculate $\chi^{\, 2}_{\, SNO}$. Event
rates 
for each type of reaction are 
\begin{eqnarray}\label{e13}
R^{ES}_k= R^{ES}_{i,j} &=& \phi_B\int\! dE_\nu\, \lambda_B (E_\nu)
\big[ \sigma_{\nu_e}^i(E_\nu)  \langle P_{ee}^j (E_\nu,t) \rangle
+\sigma_{\nu_x}^i(E_\nu) (1-\langle P_{ee}^j (E_\nu,t)\rangle )\big],
\\
R^{CC}_k= R^{CC}_{i,j} &=& \phi_B\int\! dE_\nu\, \lambda_B (E_\nu)
\big[ \sigma_{CC}^i(E_\nu)  \langle P_{ee}^j (E_\nu,t) \rangle
\big],
\\
R^{NC}_k= R^{NC}_{i,j} &=& \phi_B\int\! dE_\nu\, \lambda_B (E_\nu)
\big[ \sigma_{NC}^i(E_\nu)
\big]. 
\end{eqnarray}
In our calculations we used the neutrino-deuteron cross-sections of 
Ref. \cite{Butler:1999sv} calculated using the effective field 
theory approach. We fixed the counter-term, $L_{1A}$, of this 
approach so that it reproduces the calculation of Ref. 
\cite{Ying:1991tf} 
which incorporates the first-forbidden matrix elements in the 
calculation of the neutrino-deuteron cross sections.  
Theoretically expected rate is calculated by adding background
contributions 
like the so-called Low Energy Background (LB) and Neutron Background 
(NB) to
the sum of CC, NC and ES events. 
We take these background contributions from Ref. \cite{HOWTOSNO}. 
The expected event rate for each bin is 
\begin{equation}\label{e14a}
R^{th}_k=R^{ES}_k+R^{CC}_k+R^{NC}_k+R^{LB}_k+R^{NB}_k
\end{equation}

In calculating error matrix for SNO zenith-spectrum bins, experimental
rates and  uncertainties are taken from Ref. \cite{HOWTOSNO}.  
Statistical 
errors are calculated from the data reported by SNO and systematic
uncertainties (shape scale and resolution) are taken from
Ref. \cite{Fogli:2002pt}. Other systematics like vertex accuracy,
neutron-capture efficiency, etc. are taken from
Ref. \cite{Ahmad:2002jz}.  
In calculating $\chi^{\,  2}_{\rm \,SNO}$ we multiply sum of CC, NC
and ES 
events (without backgrounds)  by a free normalization parameter 
``$\eta$'' and minimize $\chi^{\, 2}_{\rm \,SNO}$ with respect to
$\eta$ as we did for SK (cf. Eq. (\ref{e9}). In this manner total
$^8B$ flux allowed to float freely without affecting backgrounds. 

The last component of $\chi^{\, 2}_{\,\odot}$ is from the
radiochemical experiments. In the evaluation of error
matrix for $\chi^{\, 2}_{\rm \, Cl, Ga \> Rates}$ we follow the
procedure  described in Ref. \cite{Fogli:1999zg}. 

In the presence of oscillations,  energy averaged cross section $C^{
th}_{m i}$ of neutrinos  from source $m$ at detector $i$ can be
calculated by convolving  the neutrino spectrum $\lambda_m(E_\nu)$ for
the corresponding source, the cross section  $\sigma_i(E_\nu)$ for
the corresponding detector and the survival probability $\langle
P (\nu_e \rightarrow \nu_e, E_\nu) \rangle$ (averaged over source
distributions in the Sun 
and weighted with the exposure function of each detector)
\begin{equation}
\label{e14}
C^{ th}_{m i} = \int\!  dE_\nu \, \lambda_m(E_\nu) \,
\sigma_i(E_\nu)   \, \langle P (\nu_e \rightarrow \nu_e, E_\nu)
\rangle. 
\end{equation}
Then event rate at  detector $i$ is simply:
\begin{equation}
\label{e15}
R^{ th}_{i} = \sum_m R^{ th}_{mi}
= \sum_m \phi_m C^{ th}_{m i}
\end{equation}

SSM neutrino fluxes $\phi_m$ depend on the SSM input parameters
$X_k$. The correlations between neutrino fluxes are 
parameterized by the logarithmic derivative: 
\begin{equation}
\label{e16}
\alpha_{\, i k}=\frac{\partial\ln \phi_{\, i}^{\, SSM}}{\partial\ln
  X_{\, k}} 
\end{equation}
$\Delta \ln X_{\, k}$ and $\Delta \ln C_{\, k i}$ are $1\sigma$ 
relative errors of   
SSM input parameters and energy averaged 
cross sections respectively. We adopt the values of these parameters
from Ref. \cite{Fogli:1999zg}.

From all above, one can write
\begin{equation}
\label{e17}
\sigma^{\,2}_{\,ij} = \delta_{\,ij} \, \sigma_{\, i}^{\, exp}
\sigma_{\,  
j}^{\, exp}+ \delta_{\,ij} \sum_k R^{\, thr}_{\, ki} R^{\, thr}_{\,
kj}\,(\Delta \ln C_{\,  
ki})^{\,2} + \sum_{mn} R_{\, mi}^{\, thr}\, R_{\, nj}^{\, thr}
\sum_k \alpha_{\, mk}\,\alpha_{\, nk}\,(\Delta \ln X_{\, k})^{\, 2}\ .
\end{equation}

For global analysis we take: 
\begin{equation}\label{e18}
\chi^2_{\rm Global}=\chi^2_{\,\odot}+\chi^2_{\rm KamLAND}
\end{equation}
KamLAND  detects reactor neutrinos in 1 kiloton of liquid scintillator
through the reaction: 
\begin{equation}\label{e19}
p + \overline{\nu}_{e} \rightarrow n + e^{+}.
\end{equation}
We use the phenomenological parameterization of the energy spectrum of
the incoming antineutrinos given in Ref. \cite{Vogel:iv}:
\begin{equation}
\frac{{\rm d}N_{\bar{\nu}_e}}{{\rm d}E_{\nu}}
\propto e^{a_0+a_1E_{\nu}+a_2E^2_{\nu}},
\label{eq:spectrum}
\end{equation}
where the parameters vary with different isotopes.  The spectra of
antineutrinos coming from each detector, $\phi_i (t, E_{\nu})$, can be
calculated using 
the thermal power and the isotropic composition of each 
detector. 
The effects of incomplete knowledge of the fuel composition are
explored in Ref. \cite{Murayama:2000iq}. We use the
time-averaged fuel composition for the nuclear reactors given by the 
KamLAND collaboration \cite{:2002dm}. 

Survival probability for electron
antineutrinos coming from the reactor $i$ is: 
\begin{eqnarray}
P({\overline \nu}_e \rightarrow {\overline \nu}_e) &= &
\sin^4\theta_{13} + \cos^4\theta_{13}
\left[ 1 - \sum_i \sin^22\theta_{12}\sin^2\left(\frac{1.27\Delta
m^2_{21}L_i}{E_{\nu}}\right)\right],
\label{prob}
\end{eqnarray}
where $L_i$ are the reactor-detector distances. We denote 
the energy resolution function of KamLAND by $R(T_, T')$ where $T, T'$ 
are the observed and the true positron energies. The energy resolution
is given as 
$7.5\%/\sqrt{E({\rm MeV})}$ \cite{:2002dm}. 
The number of expected events for each bin at KamLAND can be
calculated by convolving the cross section, $\sigma(E_{\nu})$, with
the reactor spectra, survival probabilities and the resolution
function of KamLAND: 
\begin{equation}\label{e20}
N_i^{\rm th} =
\int dE_\nu  \sigma(E_\nu)
\sum_j  \frac{\phi_j (t, E_{\nu})}{L_j^2} 
P_j({\overline \nu}_e \rightarrow {\overline \nu}_e, E_\nu)
\int dT R(T,T')\, ,
\end{equation}
where the electron kinetic energy is $T' \sim E_{\nu} - 0.8$ MeV, and 
$\sigma(E_{\nu})$ is the 
lowest order cross section given in
Refs. \cite{Vogel:1983hi,Vogel:1999zy}: 
\begin{equation}
\sigma(E_{\nu})=\frac{2 \pi^2}{m_e^5 f \tau_{n}}p_{e} E_{e} 
\end{equation} 
in which $E_e = T' + m_e$, 
$f=1.69$ is the integrated Fermi function for neutron
beta decay, and $\tau_n$ is the neutron lifetime. 

KamLAND reports its results in 13
bins above the threshold. Due to low statistics, we use the
prescription 
of \cite{Hagiwara:fs} in analyzing KamLAND data. We write 
\begin{equation}
\label{son}
\chi^2_{\, KamLAND-Spect}=2 \sum_i \left[ (\eta N_i^{\, th}-
N_i^{\, exp})+ N_i^{\, exp} \ln\left(\frac{N_i^{\, exp}}{\eta
N_i^{\, th} }\right)\right] +
\frac{(\eta - 1)^2}{\sigma_{\, sys}^2} ,
\end{equation}
where $\sigma_{\, sys}^2=6.75 \%$ is the systematic uncertainty, and
minimize the sum in Eq. (\ref{son}) with respect to $\eta$. 
For the total rate analysis we use  
\begin{equation}\label{e21}
\chi^2_{\, KamLAND-Rate}= \frac{( N^{\, th}-N^{\, exp} )^2  }
{\sigma_{\rm 
R}^2}\;,
\end{equation}
with errors added in the quadrature to calculate $\sigma_{\rm R}^2$.  
$N^{\, exp}$ is given
in \cite{:2002dm} and $N^{\, th}$ is calculated at each value of
oscillation parameters similar to 
the binned expected event rates.

%----------------------------------------------------------------------------%
\section{Results and Conclusions}
%----------------------------------------------------------------------------%
In our calculations we use the neutrino spectra given by 
the Standard Solar Model of Bahcall and collaborators
\cite{Bahcall:2000nu}. It it is numerically more convenient to follow 
the evolution of the matter eigenstates in the Sun and in the Earth
(or 
mass eigenstates in vacuum), a procedure which we adopted.
We take into account the distribution of
various neutrino sources in the core of the Sun and resulting
non-linear paths of the neutrinos. Thus neutrinos coming
from the other side of the Sun may have double resonances. In the Sun
we used the Landau-Zener approximation
\cite{Balantekin:1996ag,Haxton:dm,Parke:1986jy}.  
We divided the Sun into several shells which were in turn divided into 
several angular bins, calculated the derivative of
the electron density in the Landau-Zener approximation numerically for
both radial and non-radial neutrino paths and averaged the survival
probabilities over the initial source distributions. 

Survival probabilities in the Earth depend on zenith angles. As
was described in the previous Section we adopted the Preliminary Earth
Model density \cite{Dziewonski:xy} to solve neutrino evolution
oscillations numerically in the Earth. 

In our calculations we ignore the possibility of density fluctuations
in the Sun \cite{Loreti:1994ry,Balantekin:1996pp}. Recent data
indicate that such fluctuations are less than one percent of the
Standard Solar Model density \cite{Balantekin:2001dx}. We similarly
ignore possible mixing of sterile components. 
Sterile
neutrinos can play a very important role in supernova r-process 
\cite{McLaughlin:1999pd,Caldwell:1999zk,Fetter:2002xx} or the big-bang
nucleosynthesis \cite{Abazajian:2001vt}. It is worth emphasizing that
active-sterile mixing can be too small to be detectable in solar 
neutrino experiments (for a discussion of various possibilities see 
e.g. Ref. \cite{Bahcall:2002zh}) and yet 
may still have significant astrophysical impact. 

\begin{figure}
\includegraphics[scale=.85, bb= 100 100 510 750]{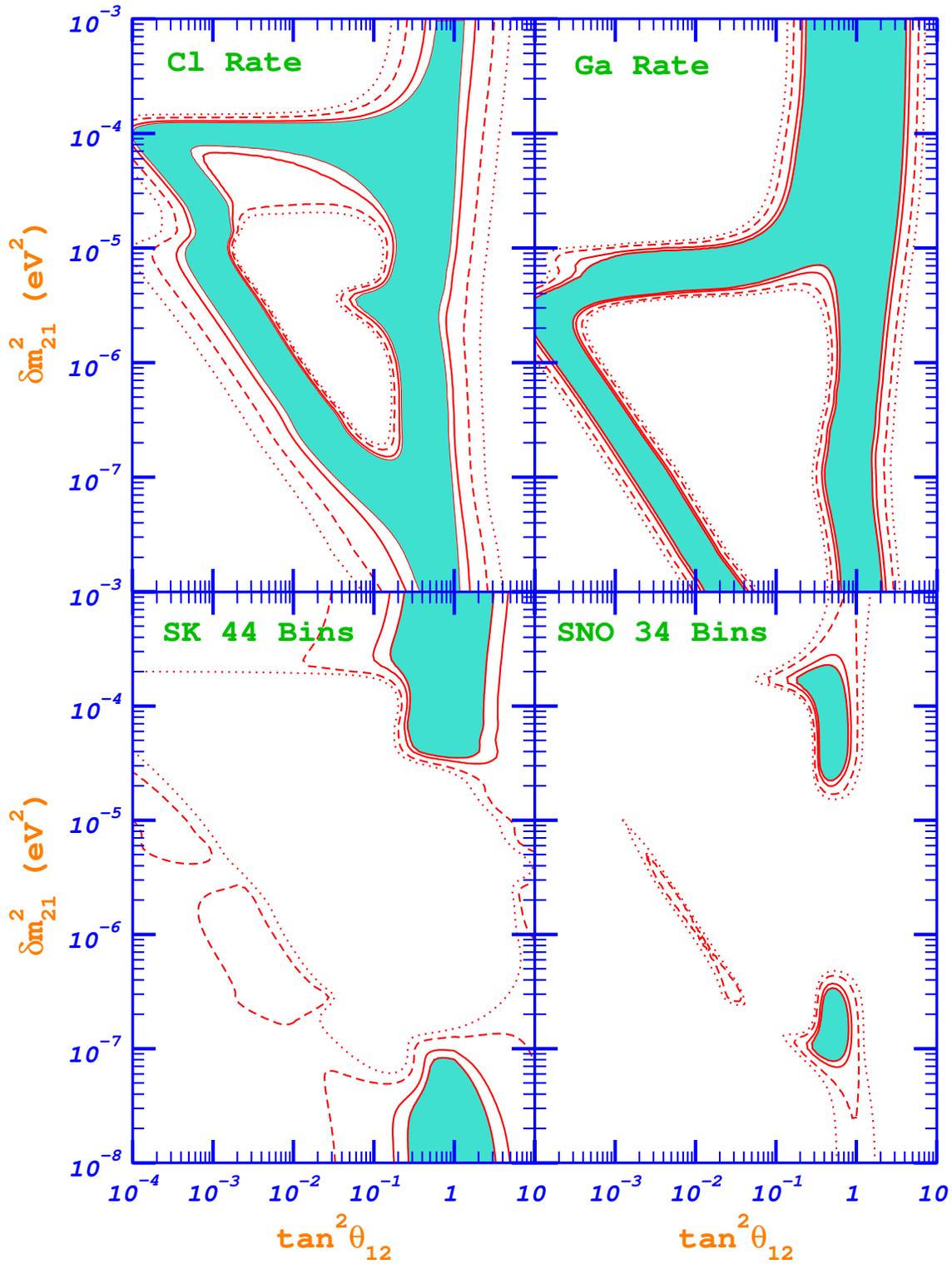}
\vspace*{+2cm} \caption{ \label{fig:2}
  Allowed regions of the neutrino parameter space when each solar
  neutrino experiment is considered separately. In this figure
  $\theta_{13}$ is taken to be zero. The shaded areas are the 90 \%
  confidence level regions. 95 \% (solid line), 99 \% (log-dashed
  line), and 99.73 \% (dotted-line) confidence levels are
  also shown.}
\end{figure}

\begin{figure}
\includegraphics[scale=1., bb= 100 100 510 750]{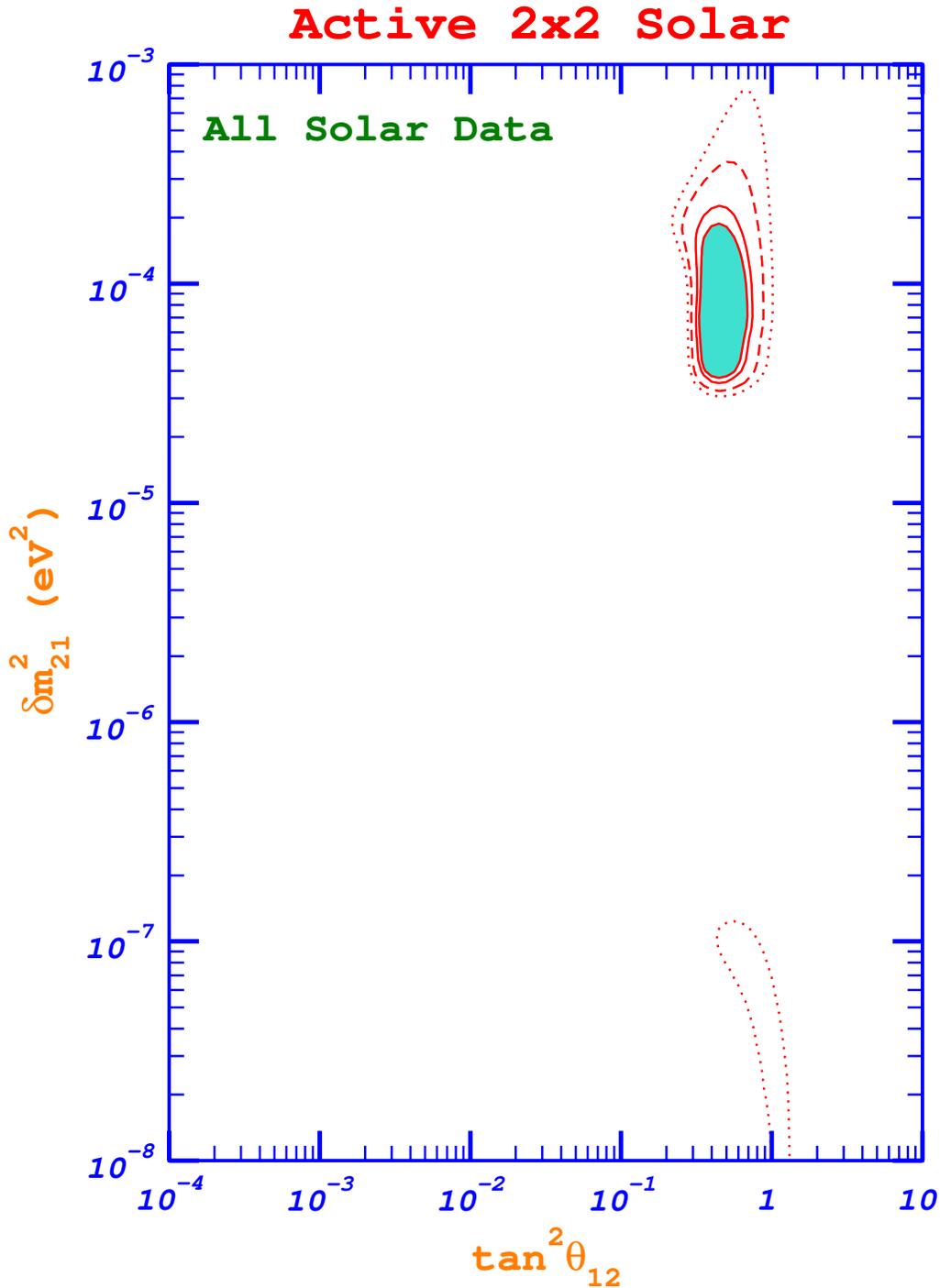}
\vspace*{-2cm} \caption{ \label{fig:3} 
  Allowed regions of the neutrino parameter space when all solar
  neutrino experiments (chlorine, all three gallium, SNO and SK
  experiments) are included in the analysis. In this figure
  $\theta_{13}$ is taken to be zero. The shaded area is the 90 \%
  confidence level region. 95 \% (solid line), 99 \% (log-dashed
  line), and 99.73 \% (dotted-line) confidence levels are
  also shown.}
\end{figure}

We first present calculations where we took the value of $\theta_{13}$
to be zero and considered only the solar neutrino data. 
Allowed regions of neutrino parameter space when each
solar neutrino experiment is considered separately are shown in
Fig. \ref{fig:2}. 
One observes that either Sudbury Neutrino
Observatory or SuperKamiokande  
individually already significantly limit the neutrino parameter
space. Our results agree well with the own analyses of these 
experimental groups.  
Allowed regions of neutrino parameter space when all 
solar neutrino experiments are combined together are shown in
Fig. \ref{fig:3}. The combined solar neutrino data already rule out 
the so-called LOW region where $\delta m_{12}^2 < 10^{-7}$ eV$^2$ at 
the 99 \% confidence level. We find the best fit (minimum $\chi^2$)  
values of neutrino parameters to be $\tan^2 \theta_{12} \sim 0.45$ 
and $\delta m_{12}^2 \sim 7.08 \times 10^{-5}$ eV$^2$ 
from our combined analysis of all the solar neutrino data. 
Our minimum $\chi^2$ value is 67.2 for 80 data points and 2 
parameters.

\begin{figure}
\includegraphics[scale=0.85, bb= 100 100 510 750]{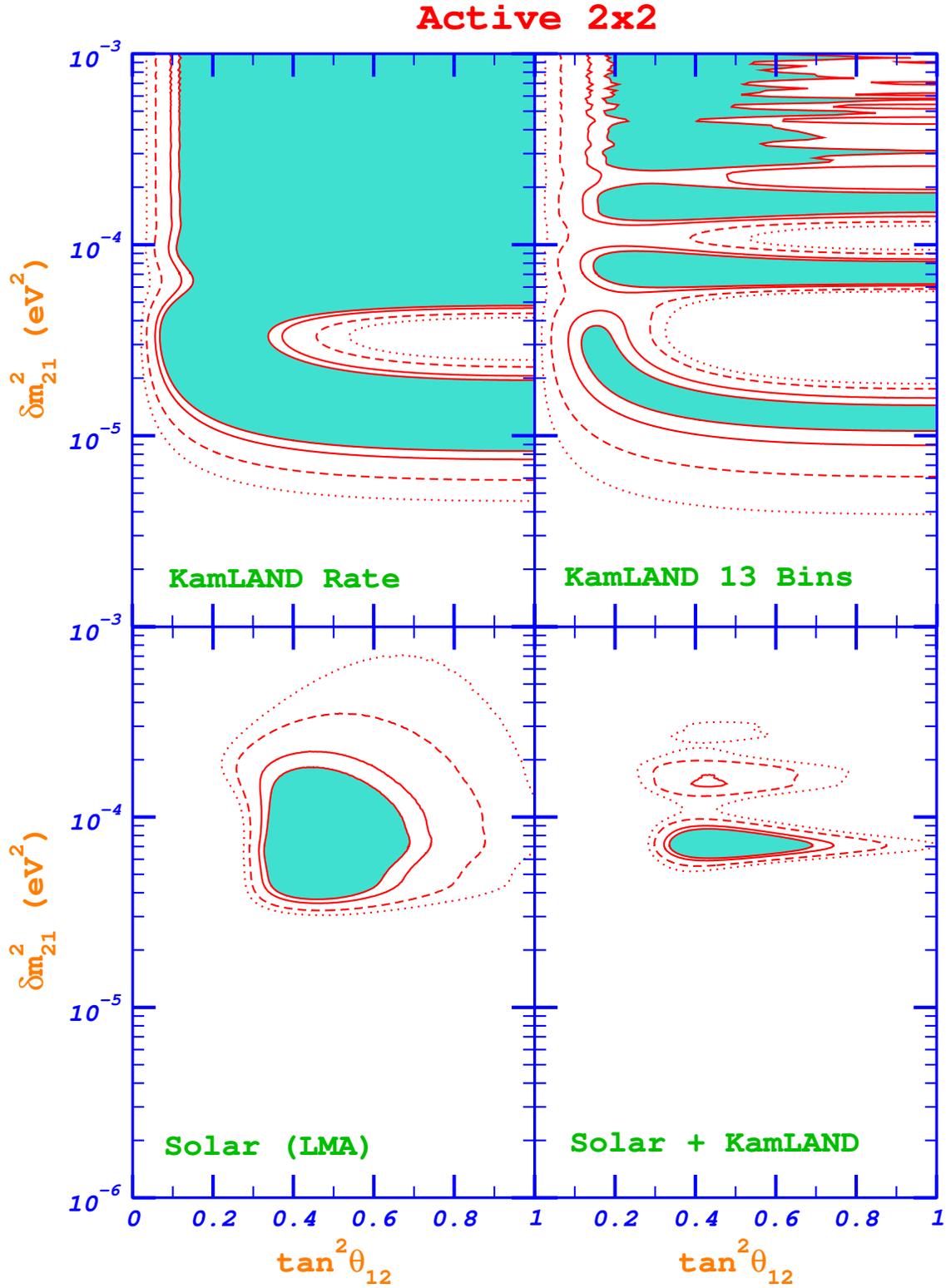}
\vspace*{+2cm} \caption{ \label{fig:4} 
  Allowed regions of the neutrino parameter space with
  $\theta_{13}$ taken to be zero. In the upper-left panel only KamLAND
  total rate and in the upper-right-hand panel binned KamLAND data are
  used. Lower-left hand panel depicts the LMA solution obtained from
  only the solar neutrino experiments. All solar neutrino experiments
  are combined with the KamLAND date to obtain the lower right-hand
  panel. The shaded areas are the 90 \%
  confidence level regions. 95 \% (solid line), 99 \% (log-dashed
  line), and 99.73 \% (dotted-line) confidence levels are
  also shown.}
\end{figure}

We next turn our attention to KamLAND data while still keeping the 
value of $\theta_{13}$ to be zero. In the upper panels of Fig. 
\ref{fig:4} we display allowed regions of the neutrino parameter 
space from the KamLAND data only. (Results with using the total rate 
only is at the left-hand side and results with using the binned data, 
which provide more information about the neutrino spectrum, is at the 
right-hand side). The result of our global analysis combining the 
solar date with data from KamLAND is shown at the lower right-hand 
side panel. For convenience of presentation we do not show the 
lower values of $\delta m_{12}^2$ on the graph, but the LOW solution 
is completely eliminated. For comparison we re-plot the neutrino 
parameter space obtained from the solar neutrino data only at the 
lower left-hand side of the Figure. KamLAND data significantly 
shrinks the LMA region (the lower right-hand side of the Figure). 
We find the best fit (minimum $\chi^2$) 
values of neutrino parameters to be $\tan^2 \theta_{12} \sim 0.46$ 
and $\delta m_{12}^2 \sim 7.1 \times 10^{-5}$ eV$^2$ 
from our combined analysis of all the solar neutrino and KamLAND 
data. 
Our minimum $\chi^2$ value is 73.2 for $\rm{80 + 13}$ data points 
and 2 parameters. 
Inclusion of the KamLAND data does not noticeably change the 
best fit values of the neutrino parameters, however KamLAND, being 
a terrestrial experiment with very different statistical errors, 
provides a completely independent test of the results from the solar 
neutrino experiments. Fig. \ref{fig:4} also illustrates that mixing 
of the (solar) neutrinos and (reactor) antineutrinos are very 
similar, very likely to be identical.  
       
\begin{figure}
\includegraphics[scale=0.85, bb= 100 100 510 750]{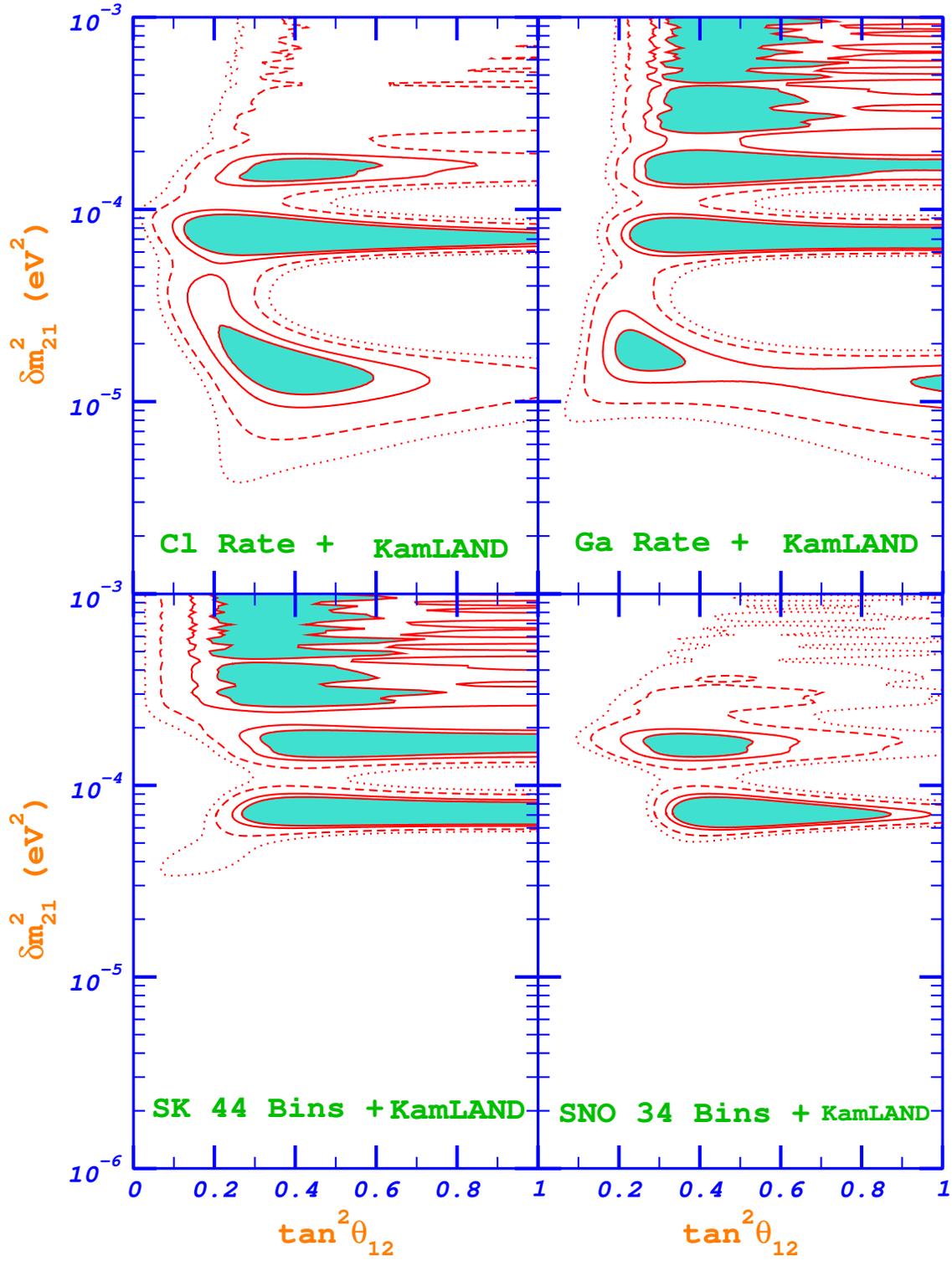}
\vspace*{+2cm} \caption{ \label{fig:5} 
  Allowed regions of the neutrino parameter space obtained when
  data from a single solar neutrino experiment is combined with
  the KamLAND data.  The shaded areas are the 90 \%
  confidence level regions. 95 \% (solid line), 99 \% (log-dashed
  line), and 99.73 \% (dotted-line) confidence levels are
  also shown.}
\end{figure}

It is instructive to investigate how well the neutrino parameter 
space is constrained. To this extend we plot allowed regions of the 
neutrino parameter spaces obtained by combining data from a single 
solar neutrino experiment with the KamLAND data in Fig. \ref{fig:5}. 
In calculating the parameter space shown in this Figure we continued 
to take the value of $\theta_{13}$ to be zero.
We observe that any single solar neutrino experiment taken together 
with KamLAND significantly constraints the parameter space. The LOW 
region, which is not shown on these plots, is again completely 
eliminated for each case. Real-time Cerenkov detectors are slightly 
more constraining than the radiochemical experiments in this regard. 
It is also interesting to realize that one no longer needs {\em all} 
the solar neutrino experiments to determine the neutrino parameters. 
We are getting closer to realizing the initial goal of the solar 
neutrino experiments, eloquently stated in the seminal papers of 
Bahcall and Davis \cite{bahcalldavis}, namely to use solar neutrino 
data to better understand the Sun. (For a preliminary effort see 
Ref. \cite{Balantekin:1997fr}).  

\begin{figure}
\includegraphics[scale=0.85, bb= 100 100 510 750]{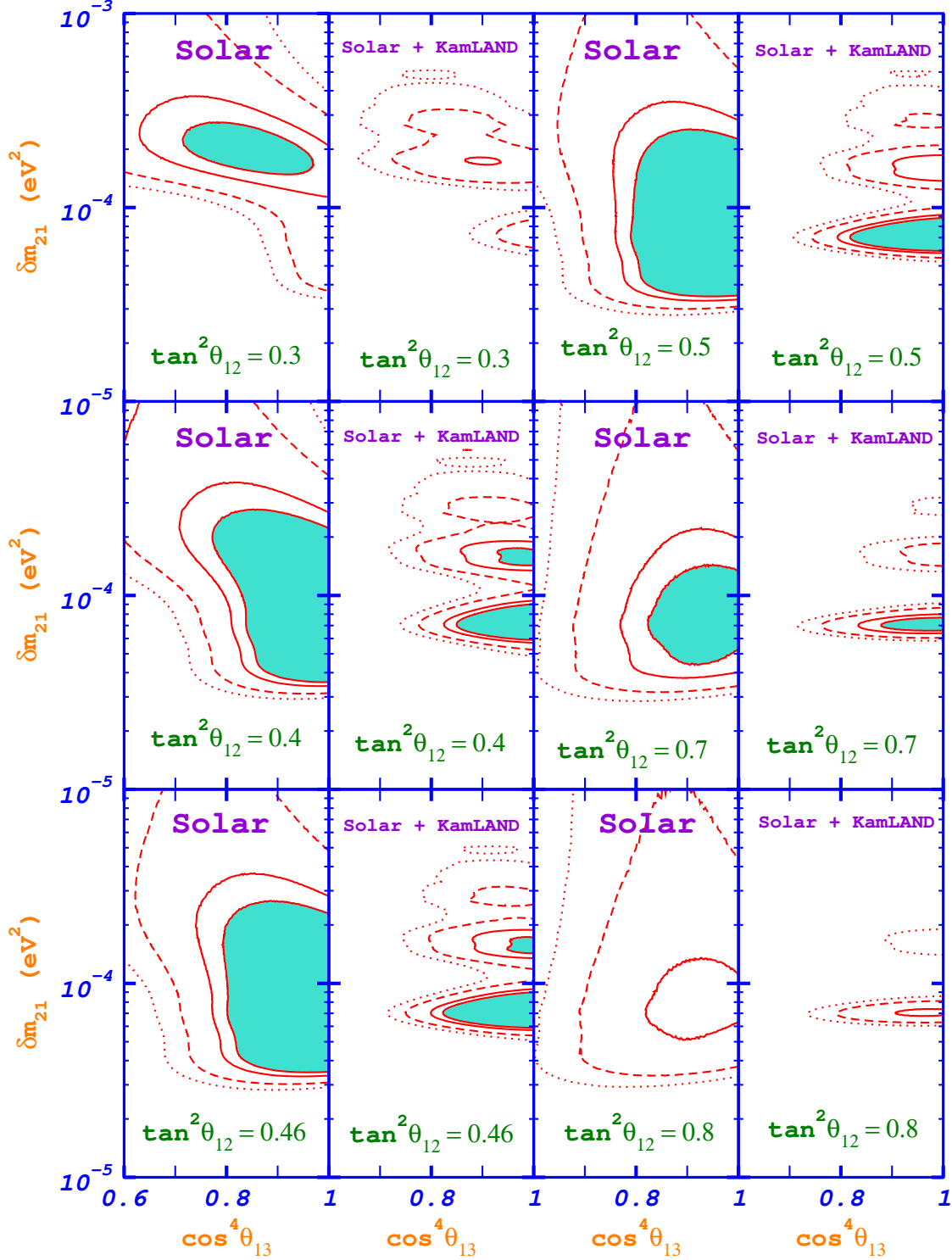}
\vspace*{+2cm} \caption{ \label{fig:6} 
  Allowed regions of neutrino parameter space when both mixing
  angles $\theta_{12}$ and $\theta_{13}$ are varied. 
  Six pairs of $\delta m_{21}^2$ vs. $\cos^4 \theta_{13}$ plots 
  are presented                          
  for representative values of
  $\tan^2 \theta_{12}$ (0.3, 0.4, 0.43, 0.46, 0.5, 0.7, and 0.8). 
  The
  figures on the left-hand side of each pair are obtained from 
  the combination of
  all solar neutrino experiments while the figures on the right-hand
  side of each pair also include the KamLAND data. The shaded areas 
  are the 90 \%
  confidence level regions. 95 \% (solid line), 99 \% (log-dashed
  line), and 99.73 \% (dotted-line) confidence levels are
  also shown.}
\end{figure}

We next examine the effects of a non-zero value of  $\theta_{13}$. 
In Fig. \ref{fig:6} we show how the parameter space changes  
as a function of $\theta_{13}$ when $\theta_{12}$ is kept fixed. 
Here we use Eq. (\ref{finalnue2}) to calculate the 3-flavor neutrino 
survival probability and perform 
the $\chi^2$ analysis for three parameters
 ($\theta_{12}$, $\theta_{13}$, and $\delta m_{12}^2$). 
We find the best fit (minimum $\chi^2$) values of neutrino 
parameters to be $\tan^2 \theta_{12} \sim 0.46$, $\cos^4 
\theta_{13} \sim 1$, and $\delta m_{12}^2 \sim 7.1 \times 10^{-5}$ 
eV$^2$. Note that the confidence level regions in the pair that 
corresponds to the best fit (the lowest right-hand side pair in 
Fig. \ref{fig:6}) are larger than the corresponding pair obtained 
with 2-flavor analysis (lower pair in  Fig. \ref{fig:4}) since as 
one goes from the former to the latter the number of parameters are 
reduced by one.
To illustrate the change in the quality of the fit as 
$\tan^2 \theta_{12}$ is changed we present confidence levels for a 
several values of $\tan^2 \theta_{12}$, some of which are clearly 
very far away from the optimal solution.  
It is interesting to note that the combined 
Solar neutrino and KamLAND date provide a limit of $\cos^4 
\theta_{13} > 0.8$ at the 90 \% confidence level. 
We are able to put such a limit since we investigated confidence 
levels for larger values of $\theta_{13}$, otherwise solar data 
alone are not sufficient to constraint this angle (cf. 
Ref \cite{Fogli:2002pb}).                           
This limit 
currently is not as good as the one obtained from the completed 
reactor 
disappearance experiments \cite{Apollonio:1999ae,Boehm:2001ik}; 
however the situation may change after a few years of data taking 
at KamLAND \cite{Gonzalez-Garcia:2001zy}.

%%%%%-----------------------------------------------------------%
\section*{ACKNOWLEDGMENTS}
%-----------------------------------------------------------------%
We thank Mark Chen and Malcolm Butler for useful conversations. 
This work was supported in part by the U.S. National Science
Foundation Grant No.\ PHY-0070161 and in part by the University of 
Wisconsin Research Committee with funds granted by the Wisconsin 
Alumni Research Foundation.

%---------------------------------------------------------------%

%----------------------------------------------------------%

%------------------------------------------------------------%
% FIGURES 
%----------------------------------------------------------------%

% \begin{figure}
% \includegraphics[scale=0.90, bb= 100 100 510 750]{graphX}
% \vspace*{+0cm} \caption{ \label{fig:X} titlex titlex titlex
% titlex  titlex titlex  titlex titlex titlex titlex titlex }
% \end{figure}

%----------------------------------------------------------------%

%---------------------------------------------------------------%

%---------------------------------------------------------------%

%----------------------------------------------------------------%

%------------------------------------------------------------%

%------------------------------------------------------------%

%----------------------------------------------------%
\end{document}